# The Exploration and Evaluation of Generating Affective 360° Panoramic VR Environments Through Neural Style Transfer

Yanheng Li[1], Long Bai[2,†], Yaxuan Mao[1,†], Xuening Peng[3], Zehao Zhang[4], Xin Tong[3*], Ray LC[1*]

## ABSTRACT

Affective virtual reality (VR) environments with varying visual style can impact users' valence and arousal responses. We applied Neural Style Transfer (NST) to generate 360° VR environments that elicited users' varied valence and arousal responses. From a user study with 30 participants, findings suggested that generative VR environments changed participants' arousal responses but not their valence levels. The generated visual features, e.g., textures and colors, also altered participants' affective perceptions. Our work contributes novel insights about how users respond to generative VR environments and provided a strategy for creating affective VR environments without altering content.

**Index Terms:** Human-centered computing—Human computer interaction—HCI design and evaluation methods; Computing methodologies—Computer graphics—Image manipulation

## 1 INTRODUCTION

The immersive environment can elicit users' affective responses by stimulating different senses, such as vision, hearing and olfaction [6, 7]. Horrible lighting, for instance, could be a passive elicitation for the sense of vision [7]. In narrative virtual reality (VR) environments, visual features can significantly elicit users' specific emotions [8]. Prior work explored how to reconstruct realistic scenes in VR to elicit users' different affective responses. For example, Felnhofer et al. [3] created five VR parks with different environments, in which each scene elicited specific affective responses, e.g., joy, sadness, boredom, anger and anxiety. However, customizing all details in VR environments is challenging, especially for eliciting specific affective responses.

Researchers have developed the neural style transfer (NST) to separate the content and style from a style image A and apply image A's style to image B's content to generate a new image with the same content as B but the same style as A [4]. Here, the style image refers to an image with a specific style reference, such as some famous painters' line strokes from existing artworks. Besides 2D images, prior work also applied NST in spherical images for 3D movies and VR [1]. Although prior work altered users' perceptions by manipulating visual elements in VR, their affective state manipulation is based on their subjective judgment. Therefore, our study aims to generate style-transferred 360° images that can affect users' affective states (valence and arousal levels) in VR. We created an affective VR generating system using the NST approach (Fig. 1) and conducted a within-subject user study to evaluate the effectiveness of our system and approach.

*Corresponding authors: Xin Tong (xin.tong@dukekunshan.edu.cn) and Ray LC(lc@raylc.org). †Co-second authors: Long Bai and Yaxuan Mao.
[1]City University of Hong Kong
[2]The Chinese University of Hong Kong
[3]Duke Kunshan University
[4]University of Waterloo

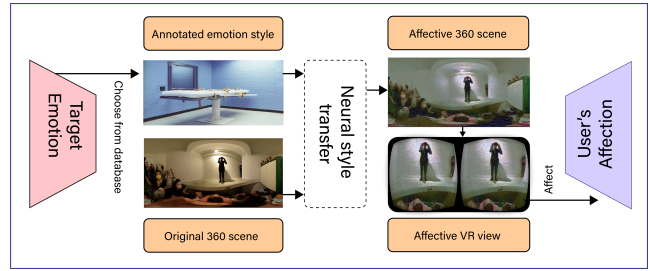

Figure 1: The affective VR generating system. First, select a style image with labeled valence and arousal values for the targeted affective outcome. Then put the style image and the 360° content image in the generator. Finally, deploy the generative affective 360° images in VR.

## 2 METHODOLOGY

### 2.1 Test Materials

We selected two content images, one for generating high- and low-valence VR conditions and the other for generating high- and low-arousal conditions (Fig. 2). Next, we deployed the pre-trained arbitrary image stylization model [1] to perform NST, and generated targeted images by applying two style images with high- and low-valence/arousal to the content image for the valence/arousal conditions, accordingly. The style images were chosen from the Geneva Affective Picture Database (GAPED)[2] [2].

### 2.2 Participants and Procedures

We recruited 30 participants in this study (10 male, 20 female, from 18 to 38 years old ($M = 22.1$, $SD = 2.28$), who tested the four 360° VR conditions (high-valence, low-valence, high-arousal, and low-arousal)in a randomly assigned order. In each VR condition, they were requested to complete an object-searching task in 4 minutes by looking through the whole environment. Precisely, participants needed to count the number of people wearing masks in the high- and low-valence conditions and find a favorite landscape in the high- and low-arousal conditions. After each condition, we conducted a brief interview and collected their ratings of the Discrete Emotion Questionnaire [5] to examine how NST-generated VR environments could impact participants' affective responses.

## 3 RESULTS

We analyzed participants' valence/arousal ratings in the high- and low- valence/arousal VR conditions to evaluate the effectiveness of generative VR environments.

### 3.1 Valence

In general, results showed that NST-generated valence conditions were likely to elicit participants' inverse valence responses due to the influence of different textures and colors obtained from style images. While 77% participants selected a negative-emotional word (like sad and angry) to describe the high-valence VR condition, some (57%) responded with a positive emotional word (like happy and relaxed) in the low-valence condition (Fig. 3). A chi-square test revealed a statistically significant difference between the high- and low-valence conditions ($p = 0.018$), which indicated that the NST approach could potentially elicit users' inverse valence levels based on the style images' valence levels. Additionally, some participants (60%) mentioned that the textures from style images slightly changed the appearance of output images compared with the content image, leading to a blurring and unfamiliar visual experience in the VR environment and negative emotions, such as fear. Also, some participants (40%) reported that brighter colors made them feel positive emotions. However, negative content like the scary human face made the bright color create tenser feelings, which could relate to individuals' emotional memory (the memory of experiences that evoked an emotional reaction).

---

[1]tfhub.dev/google/magenta/arbitrary-image-stylization-v1-256/2
[2]www4.ujaen.es/ erpadial/GAPED.html

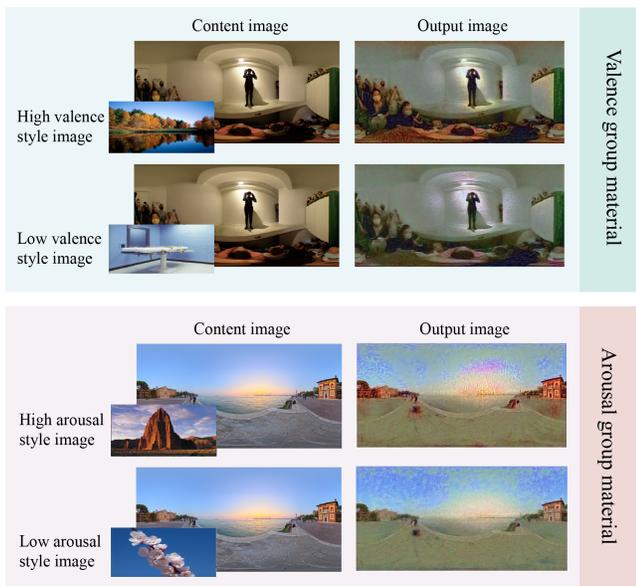

Figure 2: Image materials used for the Valence and Arousal Group.

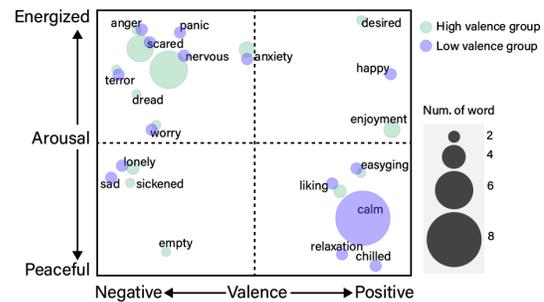

Figure 3: The bubbles' size means the frequency of each word. Most participants had negative emotions in the high-valence condition and positive emotions in the low-valence condition.

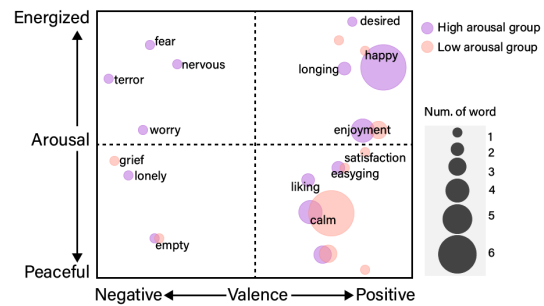

Figure 4: Most participants had high arousal levels in the high-arousal condition and low arousal levels in the low-arousal condition.

### 3.2 Arousal

Unlike findings from the two valence conditions, participants reported high-arousal affective responses when they saw the high-arousal VR condition and felt peaceful in the low-arousal condition (Fig. 4). A chi-square test revealed significant statistical differences in both test groups ($p < 0.001$). Some participants indicated that the content image in the arousal conditions was a beach landscape with sunset and fewer blurred people, which probably evoked participants' relaxed memories rather than high-arousal level (e.g., fear/scared) memories. Moreover, the high-arousal condition had a brighter color than the low-arousal condition, which made participants have improved arousal levels, according to the interviews.

### 4 CONCLUSION

Findings from our user study indicate that the VR environment generated by NST could elicit expected users' arousal states corresponding to input high/low arousal value style images. The generated texture in VR is an essential visual element that can elicit users' negative valence levels, while the color tone influences users' positive valence experience and arousal states. Thus, future work need to consider the balance of the texture and color conditions when choosing style images for altering users' valence states in NST-generated VR environments. Our system and design strategies can help VR designers quickly develop VR environments with particular affective features, such as narrative VR games and therapies for psychological treatment.